\documentclass[fleqn,usenatbib]{mnras}

\usepackage{newtxtext,newtxmath}
\usepackage[T1]{fontenc}

\DeclareRobustCommand{\VAN}[3]{#2}
\let\VANthebibliography\thebibliography
\def\thebibliography{\DeclareRobustCommand{\VAN}[3]{##3}\VANthebibliography}

\usepackage{graphicx}	% Including figure files
\usepackage{amsmath}	% Advanced maths commands
\usepackage{bm}         % Bold math fonts

\newcommand{\ketju}{\textsc{Ketju}}

\newcommand{\gadget}{\textsc{gadget-4}}

\newcommand{\Msun}{\ensuremath{M_{\sun}}}

\newcommand{\vb}[1]{\ensuremath{\bm{#1}}} %vectors

\graphicspath{{figures/}}

\title[Reviving Stochasticity]{Reviving stochasticity: uncertainty in SMBH binary eccentricity is unavoidable}

\author[A. Rawlings et al.]{
Alexander Rawlings,$^{1}$\thanks{E-mail: alexander.rawlings@helsinki.fi}
Matias Mannerkoski,$^{1}$
Peter H. Johansson,$^{1}$
Thorsten Naab$^{2}$
\vspace*{0.1cm}\\%
$^{1}$
Department of Physics,
Gustaf H\"allstr\"omin katu 2, FI-00014, University of Helsinki, Finland
\\%
$^{2}$
Max-Planck-Institut f\"ur Astrophysik, Karl-Schwarzchild-Str 1, D-85748 Garching, Germany
}

\date{Accepted XXX. Received YYY; in original form ZZZ}

\pubyear{2023}

\begin{document}
\label{firstpage}
\pagerange{\pageref{firstpage}--\pageref{lastpage}}
\maketitle

\begin{abstract}
We study supermassive black hole (SMBH) binary eccentricity of equal-mass galaxy mergers in $N$-body simulations with the \ketju{} code, which combines the \gadget{} fast multipole gravity solver with accurate regularised integration and post-Newtonian corrections around SMBHs.
In simulations with realistic, high eccentricity galactic merger orbits, the hard binary eccentricity is found to be a non-linear function of the deflection angle in the SMBH orbit during the final, nearly radial close encounter between the SMBHs before they form a bound binary.
This mapping between the deflection angle and the binary eccentricity has no apparent resolution dependence in our simulations spanning the resolution range of $1\times10^5$--$8\times10^6$ particles per galaxy.
The mapping is also captured using a simple model with an analytic potential, indicating that it is driven by the interplay between a smooth asymmetric stellar background potential and dynamical friction acting on the SMBHs.
Due to the non-linearity of this mapping, in eccentric major merger configurations small, parsec-scale variations in the merger orbit can result in binary eccentricities varying in nearly the full possible range between $e=0$ and $e=1$.
In idealised simulations, such variations are caused by finite resolution effects,
and convergence of the binary eccentricity can be achieved with increasing resolution.
However, in real galaxies, other mechanisms such as nuclear gas and substructure that perturb the merger orbit are likely to be significant enough for the binary eccentricity to be effectively random.
Our results indicate that the distribution of these effectively random eccentricities can be studied using even moderate resolution simulations.
\end{abstract}

\begin{keywords}
black hole physics -- galaxies: kinematics and dynamics -- methods: numerical -- software: simulations
\end{keywords}

%%%%%%%%%%%%%%%%%%%%%%%%%%%%%%%%%%%%%%%%%%%%%%%%%%

%%%%%%%%%%%%%%%%% BODY OF PAPER %%%%%%%%%%%%%%%%%%

% INTRODUCTION
\section{Introduction}
Supermassive black holes (SMBHs) are believed to reside at the centres of all massive galaxies \citep[e.g.][]{kormendy2013}.
In the $\Lambda$CDM model as galaxies grow through gas accretion and mergers \citep[e.g.][]{volonteri2003,naab2017} their SMBHs 
are also expected to interact in a three-phase merger process \citep{begelman1980}.

Firstly, dynamical friction \citep[DF,][]{chandrasekhar1943} acts to bring the SMBHs from kiloparsec-scales down to parsec-scale separations, after which the SMBHs form a bound binary with a semimajor axis $a$ and eccentricity $e$ \citep[e.g.][]{milosavljevic2001,merritt2013}. In the second phase, the
SMBH binary separation is reduced through sequential slingshot encounters with the surrounding stellar
distribution \citep{hills1980, hills1983, quinlan1996,rantala2018}. Finally, at small subparsec separations gravitational wave (GW) emission
becomes the dominant mechanism by which the SMBH binary can lose its remaining orbital energy and angular momentum, thus driving the
SMBHs to coalescence \citep{peters1963,peters1964}.

The complex nature of SMBH coalescence in a galaxy merger setting necessitates the use of numerical techniques \citep[e.g.][]{berentzen2009,khan2011,dosopoulou2017,mannerkoski2019}
to provide quantitative predictions for ongoing observational programmes such as ground-based pulsar timing arrays \citep[PTAs,][]{agazie2023,2023Antoniadis,2023Xu,zic2023}
and the upcoming Laser Interferometer Space Antenna \citep[LISA, e.g.][]{amaro-seoane2023}.
In particular, the eccentricity of the binary significantly affects the GW emission, with higher eccentricities resulting both in faster mergers as well as changes in the emitted GW spectrum \citep{enoki2007, huerta2015, taylor2016}.
Understanding SMBH binary eccentricity and how faithfully it is captured in numerical simulations is thus critical for predicting and interpreting observations done with instruments such as PTAs and LISA. 

The SMBH binary merging process and its dependence on eccentricity and resolution has been extensively studied using collisionless merger simulations \citep[e.g.][]{berentzen2009,vasiliev2015,bortolas2016,gualandris2017,gualandris2022}. 
Recently, \citet{nasim2020} have argued that the substantial scatter in SMBH binary eccentricity observed in gas-free merger simulations is an artefact of poor phase space
discretisation, and that in the real Universe where SMBH masses are far greater than stellar masses $(M_\bullet \gg m_\star)$, SMBH binary eccentricity is a reliably predictable quantity.   

In this paper, we find that for realistic galaxy merger orbits the scatter in the final SMBH binary eccentricity is due to the 
physical sensitivity of the final eccentricity to small perturbations of the final  nearly radial plunging trajectory of the SMBHs before they become bound, even in the infinite resolution limit.
As a result, we argue that the stochasticity of the binary eccentricity is an unavoidable physical feature of realistic galaxy mergers, at least in the equal-mass case.

This paper is structured as follows. In Section \ref{sec:num_sims} we briefly discuss the main features
of the \ketju{} code and our numerical simulations. In Section \ref{sec:results} we study the eccentricity scatter in our simulations and also present a simple model that captures the observed scatter. 
We discuss our results and their implications in Section \ref{sec:discussion} and finally present our conclusions in Section \ref{sec:conclusions}. 

\begin{figure*}
    \centering
    \includegraphics{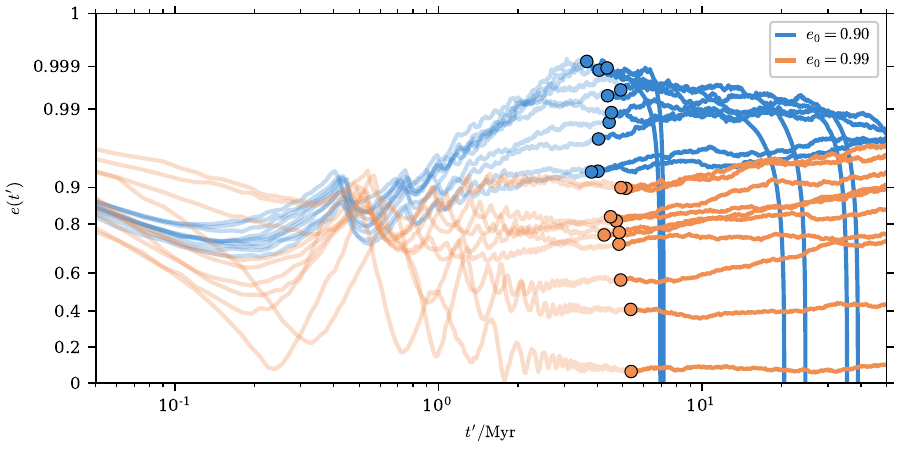}
    \caption{Binary eccentricity $e$ (note the non-linear scale) and shifted time $t'=t-t_\mathrm{bound}$ for the $M_\bullet/m_\star=5000$ resolution simulations from the $e_0 = 0.90$ set (blue lines) and $e_0 = 0.99$ set (orange lines). 
    The circle markers indicate the time when a hard binary formed, with the preceding semitransparent lines indicating the transition period when the hard binary is still forming, and the solid lines the hard binary evolution. 
    During the transition period, the Keplerian definition of binary motion is affected by the intervening stellar mass, creating the oscillatory artefacts in the semitransparent lines.
    In the $e_0=0.90$ simulation set, six simulations have a binary merger within $50\,\mathrm{Myr}$.}
    \label{fig:eccentricities}
\end{figure*}

% NUMERICAL
\section{Numerical Simulations}
\label{sec:num_sims}

We construct a number of idealised galaxy merger simulations, which we evolve with our new version of the \ketju{} code \citep{mannerkoski2023,rantala2017}.
The dynamics of
SMBHs, and stars in a small region around them, are integrated with an algorithmically regularised integrator \citep{rantala2020}, whereas the dynamics of the remaining particles
is computed with the \gadget{} \citep{springel2021} fast multiple method (FMM) with second order multipoles. Together with hierarchical time integration this allows for
symmetric interactions and manifest momentum conservation. \ketju{} also includes post-Newtonian (PN) correction terms up to order 3.5 between each pair of SMBHs \citep{blanchet2014}.

Our galaxy models represent the nuclear bulge of a gas-devoid elliptical galaxy, and are designed to match exactly the models of \citet{nasim2020}.
Each galaxy consists of a stellar-only \citet{dehnen1993} sphere with shape parameter $\gamma=0.5$ and scale radius $a=186\,\mathrm{pc}$, where the Dehnen profile is given by:
\begin{equation}\label{eq:dehnen}
    \rho_\star(r) = \frac{(3-\gamma)M_\star}{4\pi} \frac{a}{r^\gamma (r+a)^{(4-\gamma)}}.
\end{equation}
The total stellar mass is $M_\star=10^{10}\,\Msun$, and at the centre of the model galaxy a SMBH of mass $M_\bullet=10^8\,\Msun$ is placed\footnote{We set the length and mass scales of the system to match one of those presented in \citet{gualandris2022}. However, as the performed simulations are gravity-only, the system may be transformed to other mass and length scales with the same $M_\mathrm{tot}/a$ ratio without affecting the results.}.
Our galaxies lie on the observed $M_\bullet\text{--}\sigma$ relation presented in \citet{vandenbosch2016}.
We test seven different mass resolutions 
with a varying number of stellar particles: $N_\star = \{1.0\times10^5,2.5\times10^5, 5.0\times10^5, 1.0\times10^6, 2.0\times10^6, 4.0\times10^6, 8.0\times10^6\}$, corresponding to $M_\bullet/m_\star= 1000 \textnormal{--} 80000$.

We then construct isolated merger initial conditions by placing two galaxies on two different elliptical orbits with eccentricities of $e_0=0.90$ and $e_0=0.99$, at a fixed
initial separation of $D=3.72\,\mathrm{kpc}$ and a semimajor axis of $a_0=2.79\,\mathrm{kpc}$.
The $e_0=0.90$ merger orbit matches that used by \citet{nasim2020}, with the radial and tangential velocities being consistent with the values reported in \citet{gualandris2022}.
For each orbital configuration, we run ten realisations for each mass resolution, to account for stochasticity caused by the discretised phase space. 
Interactions between stellar particles are softened with a softening length of $\varepsilon = 2.5\,\mathrm{pc}$, and the \ketju{} region radius is set to $r_{\rm ketju}=3\varepsilon = 7.5\,\mathrm{pc}$.

% RESULTS
\section{Results}\label{sec:results}

\begin{figure}
    \includegraphics{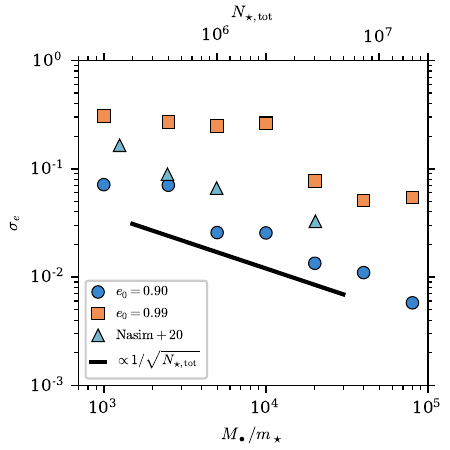}
    \caption{Convergence of the eccentricity scatter $\sigma_e$ as a function of mass resolution $M_\bullet/m_\star$ and the total number of stellar particles $N_{\star, \mathrm{tot}}$. 
    The results from \citet{nasim2020} are shown as teal triangles for comparison.
    The expected $1/\sqrt{N_\star}$ scaling is recovered for the $e_0 = 0.90$ mergers. 
    For the $e_0=0.99$ mergers, $\sigma_e$ does not decrease until $M_\bullet/m_\star \geq 20000$.}
    \label{fig:convergence}
\end{figure}

\subsection{Eccentricity scatter in simulations}

The SMBH binary eccentricity for both the $e_0=0.90$ (blue) and $e_0=0.99$ (orange) orbits are shown in \autoref{fig:eccentricities} as a function of shifted time $t' = t - t_\mathrm{bound}$, where $t_\mathrm{bound}$ is the time when the binary orbital energy $E$ becomes permanently negative. 
At the time $t_\mathrm{bound}$ the SMBH binary is not yet isolated due to the large quantity of intervening stellar mass, which causes the Keplerian definition of eccentricity to oscillate and not instantaneously match exactly the actual SMBH binary orbit. In \autoref{fig:eccentricities} we indicate the time of hard binary formation by circle markers; from this point the Keplerian eccentricity appropriately describes the binary orbit.
The binary hardening radius $a_\mathrm{h}$ is defined as \citep[e.g.][]{merritt2006}:
\begin{equation}\label{eq:ahard}
    a_\mathrm{h} = \frac{q}{(1+q)^2} \frac{r_\mathrm{m}}{4} = \frac{r_\mathrm{m}}{16},
\end{equation}
where $r_\mathrm{m}=r(m<2M_{\bullet,1})$ is the influence radius and $q$ is the mass ratio between the SMBHs (for our simulations $q=1.0$).
Before a hard binary forms, both the $e_0=0.90$ and $e_0=0.99$ sets initially demonstrate an overall decrease in eccentricity (semitransparent lines in \autoref{fig:eccentricities}) until $t'\sim1\,\mathrm{Myr}$.
It should be noted that the precise value of the eccentricity in this regime is not robust due to the SMBHs orbiting in a potential still largely influenced by the stellar background, however the overall trend is consistent with the binary orbital geometry. After $t'\sim1\,\mathrm{Myr}$, some of the $e_0=0.99$ realisations continue to have a decreasing eccentricity, and all of the $e_0=0.90$ realisations have an increasing eccentricity.
We discuss the mechanism for this in \autoref{ssec:toy}.

The $e_0=0.99$ runs show a wide variation in eccentricity, where $e$ spans almost the entire domain range $e=[0,1]$.
As a test we also perform two sets of ten runs using the $10^6$ particle resolution set up, but reduce the SMBH mass to $5.0\times10^7\,\Msun$ and $1.0\times10^7\,\Msun$, and observe the same qualitative spread in eccentricity as in the $M_\bullet=10^8\,\Msun$ case.
Even though the $e_0=0.99$ runs have a higher initial merger eccentricity than the $e_0=0.90$ runs, none of the SMBH binaries in the shown $e_0=0.99$ set obtain high enough eccentricities to undergo GW-induced coalescence during the $50\,\mathrm{Myr}$ timespan.
The $e_0=0.90$ set demonstrates six SMBH-binary mergers within $50\,\mathrm{Myr}$ of forming a bound binary, seen as a rapid orbit circularisation in \autoref{fig:eccentricities}, which is captured self-consistently using \ketju{}.

To characterise the scatter in eccentricity, we determine the mean eccentricity $e_\mathrm{h}$ over five orbital periods centred on the orbit within which the SMBH binary has become hard (equation \eqref{eq:ahard}).
We characterise the inter-simulation eccentricity scatter in the mean values of $e_\mathrm{h}$ with the standard deviation, denoted as $\sigma_e$.

We show the dependence of $\sigma_e$ on mass resolution for the $e_0=0.90$ orbit in \autoref{fig:convergence} with blue circle markers, and for the $e_0=0.99$ orbit using orange square markers.

For the $e_0=0.90$ orbit, the convergence of $\sigma_e$ scales as $1/\sqrt{N_{\star,\mathrm{tot}}}$, where $N_{\star,\mathrm{tot}}$ is the total number of stellar particles in the merger.
The values of $\sigma_e$ we report are quantitatively similar to the values found by \citet{nasim2020} for the same system, as can be seen by the teal triangles in \autoref{fig:convergence}.
The variation in eccentricity does not show the same scaling for the $e_0=0.99$ orbit as the $e_0=0.90$ orbit. 
For mass resolutions $M_\bullet/m_\star \leq 10000$, the value of $\sigma_e$ is almost constant, before significantly dropping at higher mass resolutions.

\begin{figure*}
    \centering
    \includegraphics{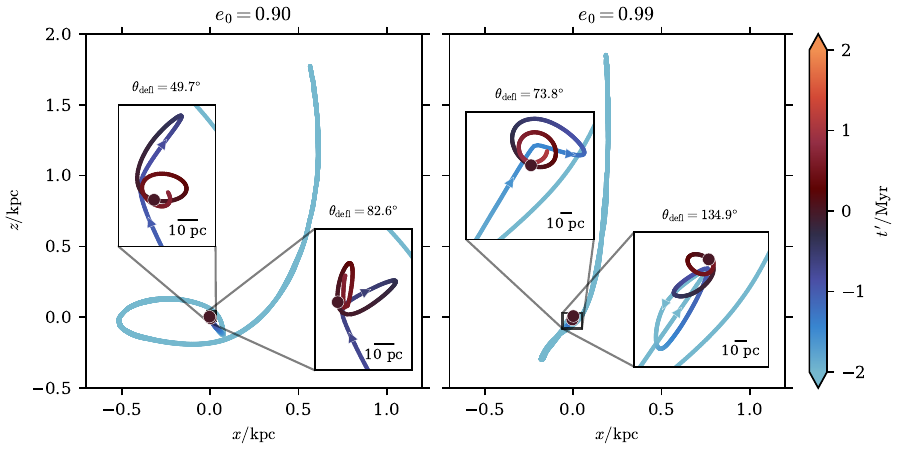}
    \caption{Orbits from $t=0\,\mathrm{Myr}$ to a time shortly after $t_\mathrm{bound}$ of a single SMBH in two representative realisations of the $e_0=0.90$ mergers (left) and the $e_0=0.99$ mergers (right), which by symmetry of the equal mass system is a reflection of the second BH orbit about $x=-z$. 
    The line gradient shows the shifted time $t'=t-t_\mathrm{bound}$, however the colouring for $|t'|>2\,\mathrm{Myr}$ is constant for visual clarity. 
    In each inset panel, $\theta_\mathrm{defl}$ is the deflection angle at the pericentre between the arrows.
    The circle marker indicates the SMBH position when a bound binary is formed.}
    \label{fig:orbit}
\end{figure*}

\subsection{Binary binding process and the scatter in eccentricity}\label{ssec:binding}

To understand the observed scatter in eccentricity, we investigate the binary binding process in each simulation. 
Before the SMBHs are bound, the primary influence on the motion is from the galactic merger potential; two realisations for both merger orbits are shown in \autoref{fig:orbit}.
The initial stages of the orbit show indiscernible variation between realisations, however after a particularly strong interaction between the SMBHs during a pericentre passage, the orbits deviate between realisations (inset panels, \autoref{fig:orbit}), and evolve to different binary eccentricities.

To quantify the strength of the hard-scattering process that randomises the SMBH orbit, we measure the effective two-body deflection angle, defined as:
\begin{equation}\label{eq:theta}
    \theta_\mathrm{defl} = 2 \arctan \left(\frac{GM}{L\sqrt{2E}} \right) = 2\arctan\left( \frac{b_{90}}{b} \right)
\end{equation}
where $M=M_{\bullet,1}+M_{\bullet,2}$, $L$ is the magnitude of the SMBH 
system angular momentum vector, $E$ the orbital energy at the time of the pericentre passage, $b$ the effective impact parameter, and $b_{90}$ the $90\degr$ deflection radius \citep{binney2008}.

\begin{figure*}
    \centering
    \includegraphics{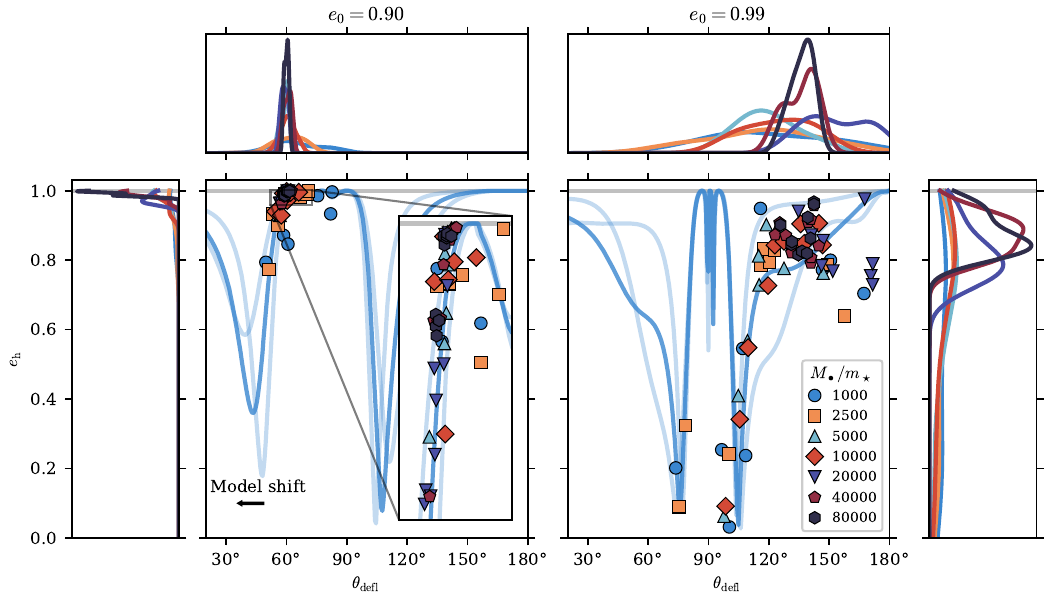}
    \caption{
    Deflection angle $\theta_\mathrm{defl}$ and resulting hard binary eccentricity
    in the $e_0=0.90$ (left) and $e_0=0.99$ (right) simulations,
    compared to fitted model curves.
    The solid lines show the results from the analytic model with $e_\mathrm{s}=0.91$,
    while the semitransparent lines show the results for  $e_\mathrm{s}=0.90$ and $e_\mathrm{s}=0.92$.
    The $e_0=0.90$ model curves have been shifted left by $14\degr$
    to better match the simulation data, with the shift indicated by an arrow.
    The marginal histograms show kernel density estimates of the simulation data.
    The inset panel for the $e_0=0.90$ data shows a zoom-in with $\theta_\mathrm{defl}=[52\degr, 73\degr]$ and $e_\mathrm{h}=[0.96, 1.00]$.
    }
    \label{fig:theta_e}
\end{figure*}

We observe clear evidence for a relationship between the deflection angle $\theta_\mathrm{defl}$ and the resulting eccentricity at the time the SMBH binary becomes hard, as shown with the data points in \autoref{fig:theta_e}.
We also observe a similar relationship in our two test simulation sets with the reduced SMBH masses.

\subsection{Reproducing the eccentricity behaviour  with a simple model}\label{ssec:toy}

The dependency of the binary eccentricity on the 
deflection angle $\theta_\mathrm{defl}$ during the binary formation
can be reproduced using a simple analytic model,
which includes only the essential components relevant to this process.
Taking the orbit to lie in the $xy$-plane,
the relative motion of the two equal-mass SMBHs in this model follows the 
equation of motion
\begin{equation} \label{eq:toy_model_eom}
\ddot{\vb{x}} = -\frac{2 G M_\bullet}{|\vb{x}|^3} \vb{x} + \vb{a}_\mathrm{bg} + \vb{a}_\mathrm{DF},
\end{equation}
where $\vb{x}=(x,y)$ is the separation vector of the SMBHs,
$M_\bullet = 10^8\,\Msun$ is the mass of a single SMBH,
$\vb{a}_\mathrm{bg}$ the acceleration due to the asymmetric background potential,
and $\vb{a}_\mathrm{DF}$ is the acceleration due to DF.

The stellar background is modelled as a constant density spheroidal potential
\citep[e.g.][]{binney2008}:
\begin{equation}
\Phi_\mathrm{bg}(\vb{x}) = \pi G \rho (A_x x^2 + A_y y^2),
\end{equation}
with the acceleration given by 
\begin{equation}
\vb{a}_\mathrm{bg}(\vb{x}) = - \nabla \Phi_\mathrm{bg}(\vb{x}).
\end{equation}
The $A$ coefficients are related to the eccentricity $e_\mathrm{s}$ of the spheroid as
\begin{align}
A_x &= 2 \left(\frac{1-e_\mathrm{s}^2}{e_\mathrm{s}^2}\right) 
    \left[\frac{1}{2 e_\mathrm{s}} \ln\left(\frac{1+e_\mathrm{s}}{1-e_\mathrm{s}}\right) - 1 \right]\\
A_y &= \frac{1-e_\mathrm{s}^2}{e_\mathrm{s}^2} 
        \left[ \frac{1}{1-e_\mathrm{s}^2} 
        - \frac{1}{2 e_\mathrm{s}} \ln\left(\frac{1+e_\mathrm{s}}{1-e_\mathrm{s}}\right)\right],
\end{align}
with the long axis aligned along the $x$-axis.
The stellar density is set to $\rho = 300 \,\Msun\,\mathrm{pc}^{-3}$,
which approximately matches the values seen in the simulations when the SMBH binary is becoming bound.

\begin{figure*}
    \centering
    \includegraphics{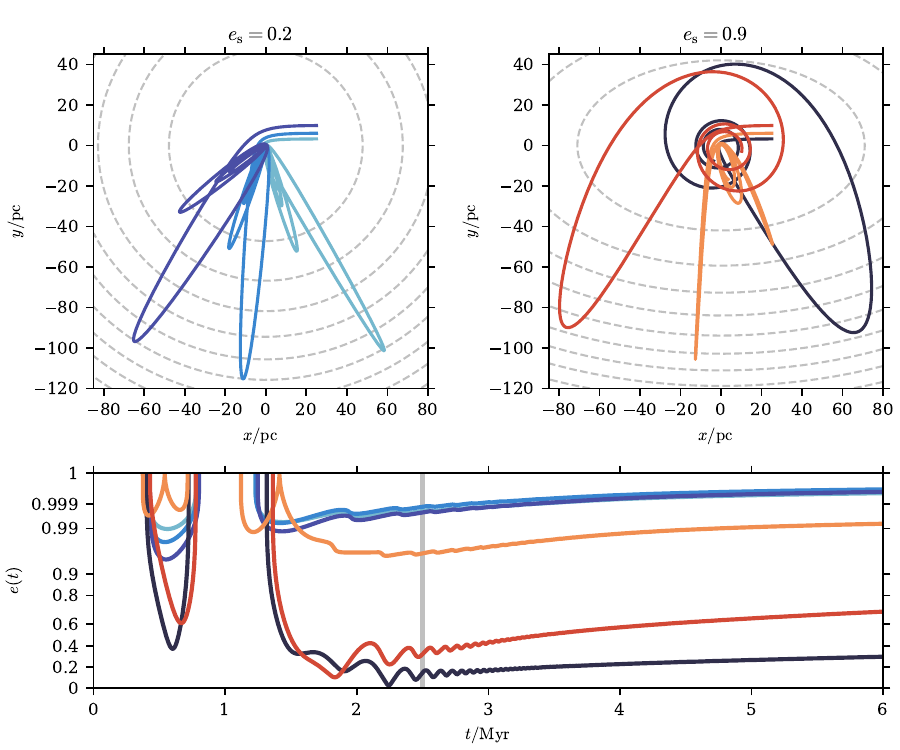}
    \caption{
    Top:
    Sample orbits computed with the analytic model for different background potential eccentricities $e_\mathrm{s}$,
    with otherwise identical initial conditions.
    The dashed lines show the isopotential contours of the model.
    Only the early part of the orbital evolution is shown for clarity.
    Bottom:
    Orbital eccentricity $e$ of the model orbits (note the non-linear scale).
    The vertical line marks the end of the period shown in the top panels. 
    }
    \label{fig:model_orbits}
\end{figure*}

The DF acting on a single BH is modelled using the \citet{chandrasekhar1943} formula assuming
a Maxwellian distribution with a constant velocity dispersion $\sigma_\star$ \citep[e.g.][]{binney2008}:
\begin{equation}\label{eq:chandra_df}
\vb{a}_{1,\mathrm{DF}} = -\frac{4 \pi G^2 M_\bullet \rho \ln{\Lambda}}{|\vb{v}_{1}|^3}
                         \left(\operatorname{erf}(X) - \frac{2X}{\sqrt{\pi}} \exp(-X^2)\right)\vb{v}_{1},
\end{equation}
where $\vb{v}_1 = \dot{\vb{x}}/2$ is the velocity of a single BH,
and $X = |\vb{v}_1|/(\sqrt{2}\sigma_\star)$.
We set ${\sigma_\star=200\,\mathrm{km\,s^{-1}}}$ based on the value measured from the simulations.
The value of the Coulomb logarithm is expected to be in the range $\ln{\Lambda}\sim 4\textnormal{--}5$
based on the size of the stellar system,
and we find that $\ln{\Lambda} = 4.7$ gives results that agree well with the simulations.

The effect of DF weakens as the SMBH binary orbit shrinks.
To account for this, when the binary is bound 
we multiply the acceleration given by equation~\eqref{eq:chandra_df}
with the smooth cut-off function
\begin{equation}
f(a) = \frac{1}{1 + \exp[(a_\mathrm{c}-a)/d_\mathrm{c}]}.
\end{equation}
Here $a$ is the semimajor axis of a bound binary,
and the cut-off scales are $a_\mathrm{c}=2 a_\mathrm{h}$
and $d_\mathrm{c} = 0.5 a_\mathrm{h}$,
with $a_\mathrm{h}$ defined by equation~\eqref{eq:ahard}.
The total DF term is then
\begin{equation}
\vb{a}_{\mathrm{DF}} = 2 f(a) \vb{a}_{1,\mathrm{DF}}.
\end{equation}

To mimic the nearly linearly plunging orbits before the scattering event seen in \autoref{fig:orbit},
we specify the initial conditions as 
\begin{align}
\vb{x}_0 &= (25\,\mathrm{pc}, b)\\
\dot{\vb{x}}_0 &= (-v_0, 0).
\end{align}
In order to match the typical SMBH relative velocity seen at this separation in
the simulations we set $v_0 = 450\,\mathrm{km\,s^{-1}}$ for the $e_0=0.90$ case, and
 $v_0 = 560\,\mathrm{km\,s^{-1}}$ for the the $e_0=0.99$ case.
 
The spheroid eccentricity parameter $e_\mathrm{s}$ is not easily measured from the simulations,
due to the presence of stellar components that remain tightly bound to the SMBHs and obscure the background
potential relevant for the dynamics,
in combination with a constantly evolving potential.
However, only sufficiently large values of $e_\mathrm{s}$ allow for low binary
eccentricities to be produced,
as is shown by \autoref{fig:model_orbits}.
Performing calculations with different values of $e_\mathrm{s}$,
we found a good fit to the simulation results for $e_\mathrm{s}\approx 0.9$.

We compute the resulting eccentricities for a range of impact parameters up to $b=20\,\mathrm{pc}$
for the two initial velocities,
by solving the equation of motion \eqref{eq:toy_model_eom} until the binary has become hard using
the error controlled 8th order Runge--Kutta method DOP853 included in the SciPy library \citep{virtanen2020}.
For reference, the $90\degr$ deflection radii are $b_{90}\approx 5.8\,\mathrm{pc}$ and
$b_{90}\approx 3.3\,\mathrm{pc}$ for the $e_0=0.90$ and $e_0=0.99$ cases, respectively.

The resulting model curves are shown together with the simulation data in \autoref{fig:theta_e}.
To correctly match the data in the $e_0=0.90$ case, the model curve has been shifted to the left by $14\degr$.
This shift is required likely due to the rotation of the stellar component tilting the background potential
relative to the SMBH trajectories in the simulations, with the effect being only evident in the $e_0=0.90$ case
due to the less radial merger orbit.
In the $e_0=0.99$ case we can also see that the simulation data has significant scatter around
the model curve in the $\theta_\mathrm{defl} \gtrsim 120\degr$ region,
although the simple model does appear to capture the mean behaviour  relatively well, apart from the very largest values of
$\theta_\mathrm{defl}$.
The behaviour of the model in this region is also sensitive to the exact values of $e_\mathrm{s}$
and $\ln{\Lambda}$,
which might also be related to the large scatter seen in the simulation data.

However, in general the behaviour  of the simulation data is captured well,
with the model correctly producing the two main eccentricity minima,
as well as the $e\approx 1$ region between them.
The predicted minima occur for trajectories where the torque from the background potential together with the DF
causes the SMBHs to loop around into a nearly circular orbit, resulting in the  low eccentricities seen in some of the $e_0=0.99$ simulation data.
Conversely, highly eccentric binaries are produced when the binary is trapped into nearly radial oscillations along 
the main axes of the potential.
These mechanisms are contrasted in \autoref{fig:model_orbits}.

% DISCUSSION
\section{Discussion}
\label{sec:discussion}

As shown by both the simulations and the analytic model, the eccentricity of the hard SMBH binary in the studied merger configurations can span nearly the full possible range in the interval $e_\mathrm{h}=[0,1]$, depending on comparatively small, parsec-scale differences in the particular realisation of the galactic merger orbit. On the other hand, in minor mergers where the stellar background is less asymmetric when the SMBHs become bound,
the scatter in eccentricity is likely to be relatively low even in the case of a radial merger orbit, since the system is less sensitive to the exact value of the deflection angle,
as seen in the $e_\mathrm{s}=0.2$ case in \autoref{fig:model_orbits}.

Merger orbits with a lower initial eccentricity $e_0$ than the values in this work can be expected to show less scatter in the binary eccentricity due to the lack of a hard scattering event that is sensitive to slight perturbations in the merger orbit, which was also seen in the study by \citet{gualandris2022}.
However, major mergers are expected to occur on highly radial orbits based on cosmological simulations \citep[e.g.][]{khochfar2006}. In addition, the DF from the dark matter halo would in general drive even lower eccentricity merger orbits to highly radial final plunges when the initial separation of the galaxies is large enough.

The simulation setup used in this work is idealised, and covers only a small part of the parameter space of merger configurations.
However, similar behaviour  can be expected to occur also in more realistic models,
since the relevant dynamics occurs within a few hundred parsecs from the centre of the merged galaxy,
where we do not expect any significant effects from dark matter or the outer parts of the galaxy, which are not included in our present simulations.
Preliminary results using steeper $(\gamma \gtrsim 1)$, more observationally-motivated galaxy density profiles indicate that similar behaviour occurs also in more realistic merger systems. 

The variation of the deflection angle is caused by the random variation of the impact parameter $b$ through equation~\eqref{eq:theta}.
In numerical simulations with a discretised phase space, the primary cause of random variation in $b$ is the Brownian motion of the SMBHs before they form a bound binary \citep{merritt2001,bortolas2016}.
This effect scales as $1/\sqrt{N_{\star,\mathrm{tot}}}$, which can explain the observed scaling of $\sigma_e$ \citep{nasim2020} when the system falls into the region of parameter space where the relation between $e$ and $b$
is approximately linear.
The deviation from this scaling seen in \autoref{fig:convergence} can then be explained by the fact that the eccentricity is not globally a linear function of the impact parameter. 
The non-linear mapping also explains why GW-induced mergers were present within a short ($<50\,\mathrm{Myr}$) timeframe for the $e_0=0.90$ mergers and not the $e_0=0.99$ mergers in \autoref{fig:eccentricities}, and highlights the difficulty in predicting the binary eccentricity $e_\mathrm{h}$ from the initial merger orbit.

While the scatter in the impact parameter between different numerical realisations of the same merger is due to relatively low number of particles compared to real galaxies,
we expect that uncertainty of a similar magnitude is also present in real systems due to various mechanisms, such as perturbations of the SMBH motion from gas and substructure in their stellar environment, or larger-scale perturbations of the galactic merger orbit.
In addition, the SMBHs are also not necesssarily located exactly at the centre of mass of the galaxy.
For example, the work by \citet{batcheldor2010} has suggested that the SMBH in M87 may be displaced from the galactic photocentre by up to $\sim7\,\mathrm{pc}$ due to jet acceleration or gravitational recoil kicks. Thus, it is not possible to give an exact prediction for the SMBH binary eccentricity produced by a given merger,
and instead we must focus on predicting the distribution of eccentricities that can be produced by a given merger configuration.

Predicting the eccentricity distribution requires the knowledge of the distribution of impact parameters that can result from essentially identical mergers
as well as the mapping between the impact parameter, or deflection angle, and the final eccentricity.
Since the relation between the impact parameter and the eccentricity does not appear to significantly depend on the resolution,
simulations at moderate mass resolutions of $M_\bullet/m_\star \sim 10^3$ can be used
to map out the relation by performing a large number of merger simulations,
possibly augmented by more sophisticated versions of the analytic model presented here.
The distribution of impact parameters in real systems is however a much more difficult problem to tackle,
since in simulations numerical resolution effects are likely dominant.

Even without such extensive studies on the full distribution of SMBH binary eccentricities, the present results suggest that the distribution is likely to be broad and contain a significant number of binaries at high eccentricities.
The possibility of a large fraction of high binary eccentricities has implications in particular for studies of the expected gravitational wave background (GWB) signal measured by PTAs, as in particular binaries with $e_\mathrm{h}\gtrsim 0.9$ can retain significant eccentricities while their GW emission is within the observable frequency band.
This could be observable in the shape of the GWB spectrum \citep{kelley2017}, and would also affect the likelihood of detecting individual SMBH binaries with PTAs \citep{kelley2018}.
It is interesting to note that \citet{bi2023} obtained a distribution of eccentricities consistent with the expectation of a significant fraction of SMBH binaries at high eccentricities based on analysis of the recent PTA observations.

In the present work, the influences of rotation and SMBH orbital flips are not considered. 
A counter-rotating stellar background with respect to the SMBH binary angular momentum has a tendency to increase the binary eccentricity, whereas a corotating stellar background tends to circularise the binary \citep{sesana2011, holley-bockelmann2015}.
The SMBH angular momentum vector can however undergo reversals as a result of gravitational torques following pericentre passages of the binary \citep[][termed an orbital flip]{rantala2019} an example of which is shown in the left inset of of the left panel of \autoref{fig:orbit}.
The chaotic reversals of the binary angular momentum act to randomise the relative rotation of the stellar background, leading in general to complex eccentricity evolution \citep{nasim2021}.
A result of a sustained co- or counter-rotating stellar background could be an evolution of the binary eccentricity after $e_\mathrm{h}$ to a bimodal distribution for a given merger configuration, if the stellar scattering interactions were strong enough.
Whilst these mechanisms are unlikely to alter much our findings concerning the hard binary eccentricity, they may obfuscate predictions for SMBH binary eccentricity prior to the merger, which is critical for GW detection missions.

In contrast to the gas-free and fairly low density systems studied here, in higher density or gas-rich systems the binary eccentricity can also evolve significantly after the binary has become hard but before it has entered the GW emission dominated regime.
In particular, interactions with a circumbinary accretion disc can lead to rapid evolution of the eccentricity if the accretion rate is high.
Studies of prograde coplanar circumbinary discs have found that at almost all eccentricities the binaries evolve towards an eccentricity of $e\approx 0.5$ \citep{zrake2021,dorazio2021}, whereas retrograde discs lead to the eccentricity always increasing \citep{tiede2023}.
However, these studies are generally limited to eccentricities below $e<0.9$.
Our results indicate the need for extending such studies to higher binary eccentricities, as well as to more general configurations such as polar discs which become increasingly likely at high eccentricities \citep[e.g.][]{martin2017}.

% CONCLUSION
\section{Conclusions}
\label{sec:conclusions}
In conclusion, we find that the variation in the hard binary eccentricity $e_\mathrm{h}$ for a given system configuration is tightly correlated with the deflection angle $\theta_\mathrm{defl}$, and thus the impact parameter $b$.
By using a simple, resolution-free analytic model of the SMBH scattering process,
we have demonstrated that uncertainty in SMBH binary eccentricity is not caused solely by discretisation effects in galaxy merger simulations,
but is rather due to the physical sensitivity of the system to small changes in the merger orbit,
which can be caused by physical mechanisms in addition to numerical discretisation effects.
The results presented here justify extending the investigation to more realistic galaxy merger scenarios,
in order to quantify the expected range of hard binary eccentricities, which is critical for predictions for current and future gravitational wave observation missions. 

% FINAL BITS
\section*{acknowledgments}
We thank the anonymous referee for their comments which contributed to improving this work.
We also thank Shihong Liao, Dimitrios Irodotou, and Francesco Paolo Rizzuto for interesting discussions.
A.R. acknowledges the support by the University of Helsinki Research Foundation.
A.R., M.M. and P.H.J. acknowledge the support
by the European Research Council via ERC Consolidator Grant KETJU (no. 818930) and the support of the Academy of Finland grant 339127.
TN acknowledges support from the Deutsche Forschungsgemeinschaft (DFG, German Research Foundation) under Germany’s Excellence Strategy - EXC-2094 - 390783311 from the DFG Cluster of Excellence ``ORIGINS''. 

The numerical simulations used computational resources provided by
the CSC -- IT centre for Science, Finland.

\section*{Author contributions}
We list here the roles and contributions of the authors according to the Contributor Roles Taxonomy (\href{https://credit.niso.org}{CRediT}). 
\textbf{AR}: Conceptualisation, Investigation, Formal analysis, Data curation, Writing -- original draft. 
\textbf{MM}: Conceptualisation, Formal analysis, Writing -- original draft. 
\textbf{PHJ}: Supervision, Writing -- review \& editing.
\textbf{TN}:  Writing -- review \& editing.

\section*{Software}
\ketju{} \citep{mannerkoski2023,rantala2017},
\gadget{} \citep{springel2021},
NumPy \citep{harris2020},
SciPy \citep{virtanen2020},
Matplotlib \citep{hunter2007},
pygad \citep{rottgers2020}.

%%%%%%%%%%%%%%%%%%%%%%%%%%%%%%%%%%%%%%%%%%%%%%%%%%
\section*{Data Availability}
The data underlying this article will be shared on reasonable request to the corresponding author.



\bibliographystyle{mnras}
\bibliography{ref} % if your bibtex file is called example.bib

\begin{thebibliography}{}
\makeatletter
\relax
\def\mn@urlcharsother{\let\do\@makeother \do\$\do\&\do\#\do\^\do\_\do\%\do\~}
\def\mn@doi{\begingroup\mn@urlcharsother \@ifnextchar [ {\mn@doi@}
  {\mn@doi@[]}}
\def\mn@doi@[#1]#2{\def\@tempa{#1}\ifx\@tempa\@empty \href
  {http://dx.doi.org/#2} {doi:#2}\else \href {http://dx.doi.org/#2} {#1}\fi
  \endgroup}
\def\mn@eprint#1#2{\mn@eprint@#1:#2::\@nil}
\def\mn@eprint@arXiv#1{\href {http://arxiv.org/abs/#1} {{\tt arXiv:#1}}}
\def\mn@eprint@dblp#1{\href {http://dblp.uni-trier.de/rec/bibtex/#1.xml}
  {dblp:#1}}
\def\mn@eprint@#1:#2:#3:#4\@nil{\def\@tempa {#1}\def\@tempb {#2}\def\@tempc
  {#3}\ifx \@tempc \@empty \let \@tempc \@tempb \let \@tempb \@tempa \fi \ifx
  \@tempb \@empty \def\@tempb {arXiv}\fi \@ifundefined
  {mn@eprint@\@tempb}{\@tempb:\@tempc}{\expandafter \expandafter \csname
  mn@eprint@\@tempb\endcsname \expandafter{\@tempc}}}

\bibitem[\protect\citeauthoryear{{Agazie} et~al.,}{{Agazie}
  et~al.}{2023}]{agazie2023}
{Agazie} G.,  et~al., 2023, \mn@doi [\apjl] {10.3847/2041-8213/acdac6}, \href
  {https://ui.adsabs.harvard.edu/abs/2023ApJ...951L...8A} {951, L8}

\bibitem[\protect\citeauthoryear{{Amaro-Seoane} et~al.,}{{Amaro-Seoane}
  et~al.}{2023}]{amaro-seoane2023}
{Amaro-Seoane} P.,  et~al., 2023, \mn@doi [Living Reviews in Relativity]
  {10.1007/s41114-022-00041-y}, \href
  {https://ui.adsabs.harvard.edu/abs/2023LRR....26....2A} {26, 2}

\bibitem[\protect\citeauthoryear{{Antoniadis} et~al.,}{{Antoniadis}
  et~al.}{2023}]{2023Antoniadis}
{Antoniadis} J.,  et~al., 2023, \mn@doi [arXiv e-prints]
  {10.48550/arXiv.2306.16227}, \href
  {https://ui.adsabs.harvard.edu/abs/2023arXiv230616227A} {p. arXiv:2306.16227}

\bibitem[\protect\citeauthoryear{Batcheldor, Robinson, Axon, Perlman  \&
  Merritt}{Batcheldor et~al.}{2010}]{batcheldor2010}
Batcheldor D.,  Robinson A.,  Axon D.~J.,  Perlman E.~S.,   Merritt D.,  2010,
  \mn@doi [\apj] {10.1088/2041-8205/717/1/L6}, 717, L6

\bibitem[\protect\citeauthoryear{Begelman, Blandford  \& Rees}{Begelman
  et~al.}{1980}]{begelman1980}
Begelman M.~C.,  Blandford R.~D.,   Rees M.~J.,  1980, \mn@doi [\nat]
  {10.1038/287307a0}, 287, 307

\bibitem[\protect\citeauthoryear{{Berentzen}, {Preto}, {Berczik}, {Merritt}  \&
  {Spurzem}}{{Berentzen} et~al.}{2009}]{berentzen2009}
{Berentzen} I.,  {Preto} M.,  {Berczik} P.,  {Merritt} D.,   {Spurzem} R.,
  2009, \mn@doi [\apj] {10.1088/0004-637X/695/1/455}, \href
  {https://ui.adsabs.harvard.edu/abs/2009ApJ...695..455B} {695, 455}

\bibitem[\protect\citeauthoryear{{Bi}, {Wu}, {Chen}  \& {Huang}}{{Bi}
  et~al.}{2023}]{bi2023}
{Bi} Y.-C.,  {Wu} Y.-M.,  {Chen} Z.-C.,   {Huang} Q.-G.,  2023, \mn@doi [arXiv
  e-prints] {10.48550/arXiv.2307.00722}, \href
  {https://ui.adsabs.harvard.edu/abs/2023arXiv230700722B} {p. arXiv:2307.00722}

\bibitem[\protect\citeauthoryear{Binney \& Tremaine}{Binney \&
  Tremaine}{2008}]{binney2008}
Binney J.,  Tremaine S.,  2008, Galactic {{Dynamics}}: {{Second Edition}}.
Princeton University Press

\bibitem[\protect\citeauthoryear{{Blanchet}}{{Blanchet}}{2014}]{blanchet2014}
{Blanchet} L.,  2014, \mn@doi [Living Reviews in Relativity]
  {10.12942/lrr-2014-2}, \href
  {https://ui.adsabs.harvard.edu/abs/2014LRR....17....2B} {17, 2}

\bibitem[\protect\citeauthoryear{{Bortolas}, {Gualandris}, {Dotti}, {Spera}  \&
  {Mapelli}}{{Bortolas} et~al.}{2016}]{bortolas2016}
{Bortolas} E.,  {Gualandris} A.,  {Dotti} M.,  {Spera} M.,   {Mapelli} M.,
  2016, \mn@doi [\mnras] {10.1093/mnras/stw1372}, \href
  {https://ui.adsabs.harvard.edu/abs/2016MNRAS.461.1023B} {461, 1023}

\bibitem[\protect\citeauthoryear{Chandrasekhar}{Chandrasekhar}{1943}]{chandrasekhar1943}
Chandrasekhar S.,  1943, \mn@doi [\apj] {10.1086/144517}, 97, 255

\bibitem[\protect\citeauthoryear{{D'Orazio} \& {Duffell}}{{D'Orazio} \&
  {Duffell}}{2021}]{dorazio2021}
{D'Orazio} D.~J.,  {Duffell} P.~C.,  2021, \mn@doi [\apjl]
  {10.3847/2041-8213/ac0621}, \href
  {https://ui.adsabs.harvard.edu/abs/2021ApJ...914L..21D} {914, L21}

\bibitem[\protect\citeauthoryear{Dehnen}{Dehnen}{1993}]{dehnen1993}
Dehnen W.,  1993, \mn@doi [\mnras] {10.1093/mnras/265.1.250}, 265, 250

\bibitem[\protect\citeauthoryear{{Dosopoulou} \& {Antonini}}{{Dosopoulou} \&
  {Antonini}}{2017}]{dosopoulou2017}
{Dosopoulou} F.,  {Antonini} F.,  2017, \mn@doi [\apj]
  {10.3847/1538-4357/aa6b58}, \href
  {https://ui.adsabs.harvard.edu/abs/2017ApJ...840...31D} {840, 31}

\bibitem[\protect\citeauthoryear{{Enoki} \& {Nagashima}}{{Enoki} \&
  {Nagashima}}{2007}]{enoki2007}
{Enoki} M.,  {Nagashima} M.,  2007, \mn@doi [Progress of Theoretical Physics]
  {10.1143/PTP.117.241}, \href
  {https://ui.adsabs.harvard.edu/abs/2007PThPh.117..241E} {117, 241}

\bibitem[\protect\citeauthoryear{Gualandris, Read, Dehnen  \&
  Bortolas}{Gualandris et~al.}{2017}]{gualandris2017}
Gualandris A.,  Read J.~I.,  Dehnen W.,   Bortolas E.,  2017, \mn@doi [\mnras]
  {10.1093/mnras/stw2528}, 464, 2301

\bibitem[\protect\citeauthoryear{Gualandris, Khan, Bortolas, Bonetti, Sesana,
  Berczik  \& {Holley-Bockelmann}}{Gualandris et~al.}{2022}]{gualandris2022}
Gualandris A.,  Khan F.~M.,  Bortolas E.,  Bonetti M.,  Sesana A.,  Berczik P.,
    {Holley-Bockelmann} K.,  2022, \mn@doi [\mnras] {10.1093/mnras/stac241},
  511, 4753

\bibitem[\protect\citeauthoryear{Harris et~al.,}{Harris
  et~al.}{2020}]{harris2020}
Harris C.~R.,  et~al., 2020, \mn@doi [\nat] {10.1038/s41586-020-2649-2}, 585,
  357

\bibitem[\protect\citeauthoryear{Hills}{Hills}{1983}]{hills1983}
Hills J.~G.,  1983, \mn@doi [\aj] {10.1086/113418}, 88, 1269

\bibitem[\protect\citeauthoryear{Hills \& Fullerton}{Hills \&
  Fullerton}{1980}]{hills1980}
Hills J.~G.,  Fullerton L.~W.,  1980, \mn@doi [\aj] {10.1086/112798}, 85, 1281

\bibitem[\protect\citeauthoryear{{Holley-Bockelmann} \&
  {Khan}}{{Holley-Bockelmann} \& {Khan}}{2015}]{holley-bockelmann2015}
{Holley-Bockelmann} K.,  {Khan} F.~M.,  2015, \mn@doi [\apj]
  {10.1088/0004-637X/810/2/139}, \href
  {https://ui.adsabs.harvard.edu/abs/2015ApJ...810..139H} {810, 139}

\bibitem[\protect\citeauthoryear{{Huerta}, {McWilliams}, {Gair}  \&
  {Taylor}}{{Huerta} et~al.}{2015}]{huerta2015}
{Huerta} E.~A.,  {McWilliams} S.~T.,  {Gair} J.~R.,   {Taylor} S.~R.,  2015,
  \mn@doi [\prd] {10.1103/PhysRevD.92.063010}, \href
  {https://ui.adsabs.harvard.edu/abs/2015PhRvD..92f3010H} {92, 063010}

\bibitem[\protect\citeauthoryear{Hunter}{Hunter}{2007}]{hunter2007}
Hunter J.~D.,  2007, \mn@doi [Computing in Science \& Engineering]
  {10.1109/MCSE.2007.55}, 9, 90

\bibitem[\protect\citeauthoryear{{Kelley}, {Blecha}, {Hernquist}, {Sesana}  \&
  {Taylor}}{{Kelley} et~al.}{2017}]{kelley2017}
{Kelley} L.~Z.,  {Blecha} L.,  {Hernquist} L.,  {Sesana} A.,   {Taylor} S.~R.,
  2017, \mn@doi [\mnras] {10.1093/mnras/stx1638}, \href
  {http://adsabs.harvard.edu/abs/2017MNRAS.471.4508K} {471, 4508}

\bibitem[\protect\citeauthoryear{{Kelley}, {Blecha}, {Hernquist}, {Sesana}  \&
  {Taylor}}{{Kelley} et~al.}{2018}]{kelley2018}
{Kelley} L.~Z.,  {Blecha} L.,  {Hernquist} L.,  {Sesana} A.,   {Taylor} S.~R.,
  2018, \mn@doi [\mnras] {10.1093/mnras/sty689}, \href
  {http://adsabs.harvard.edu/abs/2018MNRAS.477..964K} {477, 964}

\bibitem[\protect\citeauthoryear{{Khan}, {Just}  \& {Merritt}}{{Khan}
  et~al.}{2011}]{khan2011}
{Khan} F.~M.,  {Just} A.,   {Merritt} D.,  2011, \mn@doi [\apj]
  {10.1088/0004-637X/732/2/89}, \href
  {https://ui.adsabs.harvard.edu/abs/2011ApJ...732...89K} {732, 89}

\bibitem[\protect\citeauthoryear{Khochfar \& Burkert}{Khochfar \&
  Burkert}{2006}]{khochfar2006}
Khochfar S.,  Burkert A.,  2006, \mn@doi [\aap] {10.1051/0004-6361:20053241},
  445, 403

\bibitem[\protect\citeauthoryear{Kormendy \& Ho}{Kormendy \&
  Ho}{2013}]{kormendy2013}
Kormendy J.,  Ho L.~C.,  2013, \mn@doi [\araa]
  {10.1146/annurev-astro-082708-101811}, 51, 511

\bibitem[\protect\citeauthoryear{{Mannerkoski}, {Johansson}, {Pihajoki},
  {Rantala}  \& {Naab}}{{Mannerkoski} et~al.}{2019}]{mannerkoski2019}
{Mannerkoski} M.,  {Johansson} P.~H.,  {Pihajoki} P.,  {Rantala} A.,   {Naab}
  T.,  2019, \mn@doi [\apj] {10.3847/1538-4357/ab52f9}, \href
  {https://ui.adsabs.harvard.edu/abs/2019ApJ...887...35M} {887, 35}

\bibitem[\protect\citeauthoryear{{Mannerkoski}, {Rawlings}, {Johansson},
  {Naab}, {Rantala}, {Springel}, {Irodotou}  \& {Liao}}{{Mannerkoski}
  et~al.}{2023}]{mannerkoski2023}
{Mannerkoski} M.,  {Rawlings} A.,  {Johansson} P.~H.,  {Naab} T.,  {Rantala}
  A.,  {Springel} V.,  {Irodotou} D.,   {Liao} S.,  2023, \mn@doi [\mnras]
  {10.1093/mnras/stad2139}, \href
  {https://ui.adsabs.harvard.edu/abs/2023MNRAS.tmp.2061M} {}

\bibitem[\protect\citeauthoryear{{Martin} \& {Lubow}}{{Martin} \&
  {Lubow}}{2017}]{martin2017}
{Martin} R.~G.,  {Lubow} S.~H.,  2017, \mn@doi [\apjl]
  {10.3847/2041-8213/835/2/L28}, \href
  {https://ui.adsabs.harvard.edu/abs/2017ApJ...835L..28M} {835, L28}

\bibitem[\protect\citeauthoryear{Merritt}{Merritt}{2001}]{merritt2001}
Merritt D.,  2001, \mn@doi [\apj] {10.1086/321550}, 556, 245

\bibitem[\protect\citeauthoryear{Merritt}{Merritt}{2006}]{merritt2006}
Merritt D.,  2006, \mn@doi [\apj] {10.1086/506139}, 648, 976

\bibitem[\protect\citeauthoryear{Merritt}{Merritt}{2013}]{merritt2013}
Merritt D.,  2013, Dynamics and {{Evolution}} of {{Galactic Nuclei}}.
Princeton University Press

\bibitem[\protect\citeauthoryear{{Milosavljevi{\'c}} \&
  {Merritt}}{{Milosavljevi{\'c}} \& {Merritt}}{2001}]{milosavljevic2001}
{Milosavljevi{\'c}} M.,  {Merritt} D.,  2001, \mn@doi [\apj] {10.1086/323830},
  \href {https://ui.adsabs.harvard.edu/abs/2001ApJ...563...34M} {563, 34}

\bibitem[\protect\citeauthoryear{{Naab} \& {Ostriker}}{{Naab} \&
  {Ostriker}}{2017}]{naab2017}
{Naab} T.,  {Ostriker} J.~P.,  2017, \mn@doi [\araa]
  {10.1146/annurev-astro-081913-040019}, \href
  {https://ui.adsabs.harvard.edu/abs/2017ARA&A..55...59N} {55, 59}

\bibitem[\protect\citeauthoryear{Nasim, Gualandris, Read, Dehnen, Delorme  \&
  Antonini}{Nasim et~al.}{2020}]{nasim2020}
Nasim I.,  Gualandris A.,  Read J.,  Dehnen W.,  Delorme M.,   Antonini F.,
  2020, \mn@doi [\mnras] {10.1093/mnras/staa1896}, 497, 739

\bibitem[\protect\citeauthoryear{Nasim, Petrovich, Nasim, Dosopoulou  \&
  Antonini}{Nasim et~al.}{2021}]{nasim2021}
Nasim I.~T.,  Petrovich C.,  Nasim A.,  Dosopoulou F.,   Antonini F.,  2021,
  \mn@doi [\mnras] {10.1093/mnras/stab351}, 503, 498

\bibitem[\protect\citeauthoryear{Peters}{Peters}{1964}]{peters1964}
Peters P.~C.,  1964, \mn@doi [Physical Review] {10.1103/PhysRev.136.B1224},
  136, 1224

\bibitem[\protect\citeauthoryear{{Peters} \& {Mathews}}{{Peters} \&
  {Mathews}}{1963}]{peters1963}
{Peters} P.~C.,  {Mathews} J.,  1963, \mn@doi [Physical Review]
  {10.1103/PhysRev.131.435}, \href
  {https://ui.adsabs.harvard.edu/abs/1963PhRv..131..435P} {131, 435}

\bibitem[\protect\citeauthoryear{Quinlan}{Quinlan}{1996}]{quinlan1996}
Quinlan G.~D.,  1996, \mn@doi [\na] {10.1016/S1384-1076(96)00003-6}, 1, 35

\bibitem[\protect\citeauthoryear{Rantala, Pihajoki, Johansson, Naab, Lah{\'e}n
  \& Sawala}{Rantala et~al.}{2017}]{rantala2017}
Rantala A.,  Pihajoki P.,  Johansson P.~H.,  Naab T.,  Lah{\'e}n N.,   Sawala
  T.,  2017, \mn@doi [\apj] {10.3847/1538-4357/aa6d65}, 840, 53

\bibitem[\protect\citeauthoryear{{Rantala}, {Johansson}, {Naab}, {Thomas}  \&
  {Frigo}}{{Rantala} et~al.}{2018}]{rantala2018}
{Rantala} A.,  {Johansson} P.~H.,  {Naab} T.,  {Thomas} J.,   {Frigo} M.,
  2018, \mn@doi [\apj] {10.3847/1538-4357/aada47}, \href
  {https://ui.adsabs.harvard.edu/abs/2018ApJ...864..113R} {864, 113}

\bibitem[\protect\citeauthoryear{Rantala, Johansson, Naab, Thomas  \&
  Frigo}{Rantala et~al.}{2019}]{rantala2019}
Rantala A.,  Johansson P.~H.,  Naab T.,  Thomas J.,   Frigo M.,  2019, \mn@doi
  [\apj] {10.3847/2041-8213/ab04b1}, 872, L17

\bibitem[\protect\citeauthoryear{Rantala, Pihajoki, Mannerkoski, Johansson  \&
  Naab}{Rantala et~al.}{2020}]{rantala2020}
Rantala A.,  Pihajoki P.,  Mannerkoski M.,  Johansson P.~H.,   Naab T.,  2020,
  \mn@doi [\mnras] {10.1093/mnras/staa084}, 492, 4131

\bibitem[\protect\citeauthoryear{R{\"o}ttgers, Naab, Cernetic, Dav{\'e},
  Kauffmann, Borthakur  \& Foidl}{R{\"o}ttgers et~al.}{2020}]{rottgers2020}
R{\"o}ttgers B.,  Naab T.,  Cernetic M.,  Dav{\'e} R.,  Kauffmann G.,
  Borthakur S.,   Foidl H.,  2020, \mn@doi [\mnras] {10.1093/mnras/staa1490},
  496, 152

\bibitem[\protect\citeauthoryear{{Sesana}, {Gualandris}  \& {Dotti}}{{Sesana}
  et~al.}{2011}]{sesana2011}
{Sesana} A.,  {Gualandris} A.,   {Dotti} M.,  2011, \mn@doi [\mnras]
  {10.1111/j.1745-3933.2011.01073.x}, \href
  {https://ui.adsabs.harvard.edu/abs/2011MNRAS.415L..35S} {415, L35}

\bibitem[\protect\citeauthoryear{Springel, Pakmor, Zier  \& Reinecke}{Springel
  et~al.}{2021}]{springel2021}
Springel V.,  Pakmor R.,  Zier O.,   Reinecke M.,  2021, \mn@doi [\mnras]
  {10.1093/mnras/stab1855}, 506, 2871

\bibitem[\protect\citeauthoryear{{Taylor}, {Huerta}, {Gair}  \&
  {McWilliams}}{{Taylor} et~al.}{2016}]{taylor2016}
{Taylor} S.~R.,  {Huerta} E.~A.,  {Gair} J.~R.,   {McWilliams} S.~T.,  2016,
  \mn@doi [\apj] {10.3847/0004-637X/817/1/70}, \href
  {https://ui.adsabs.harvard.edu/abs/2016ApJ...817...70T} {817, 70}

\bibitem[\protect\citeauthoryear{{Tiede} \& {D'Orazio}}{{Tiede} \&
  {D'Orazio}}{2023}]{tiede2023}
{Tiede} C.,  {D'Orazio} D.~J.,  2023, \mn@doi [arXiv e-prints]
  {10.48550/arXiv.2307.03775}, \href
  {https://ui.adsabs.harvard.edu/abs/2023arXiv230703775T} {p. arXiv:2307.03775}

\bibitem[\protect\citeauthoryear{{Vasiliev}, {Antonini}  \&
  {Merritt}}{{Vasiliev} et~al.}{2015}]{vasiliev2015}
{Vasiliev} E.,  {Antonini} F.,   {Merritt} D.,  2015, \mn@doi [\apj]
  {10.1088/0004-637X/810/1/49}, \href
  {https://ui.adsabs.harvard.edu/abs/2015ApJ...810...49V} {810, 49}

\bibitem[\protect\citeauthoryear{Virtanen et~al.,}{Virtanen
  et~al.}{2020}]{virtanen2020}
Virtanen P.,  et~al., 2020, \mn@doi [Nature Methods]
  {10.1038/s41592-019-0686-2}, 17, 261

\bibitem[\protect\citeauthoryear{{Volonteri}, {Haardt}  \& {Madau}}{{Volonteri}
  et~al.}{2003}]{volonteri2003}
{Volonteri} M.,  {Haardt} F.,   {Madau} P.,  2003, \mn@doi [\apj]
  {10.1086/344675}, \href
  {https://ui.adsabs.harvard.edu/abs/2003ApJ...582..559V} {582, 559}

\bibitem[\protect\citeauthoryear{{Xu} et~al.,}{{Xu} et~al.}{2023}]{2023Xu}
{Xu} H.,  et~al., 2023, \mn@doi [Research in Astronomy and Astrophysics]
  {10.1088/1674-4527/acdfa5}, \href
  {https://ui.adsabs.harvard.edu/abs/2023RAA....23g5024X} {23, 075024}

\bibitem[\protect\citeauthoryear{{Zic} et~al.,}{{Zic} et~al.}{2023}]{zic2023}
{Zic} A.,  et~al., 2023, \mn@doi [arXiv e-prints] {10.48550/arXiv.2306.16230},
  \href {https://ui.adsabs.harvard.edu/abs/2023arXiv230616230Z} {p.
  arXiv:2306.16230}

\bibitem[\protect\citeauthoryear{{Zrake}, {Tiede}, {MacFadyen}  \&
  {Haiman}}{{Zrake} et~al.}{2021}]{zrake2021}
{Zrake} J.,  {Tiede} C.,  {MacFadyen} A.,   {Haiman} Z.,  2021, \mn@doi [\apjl]
  {10.3847/2041-8213/abdd1c}, \href
  {https://ui.adsabs.harvard.edu/abs/2021ApJ...909L..13Z} {909, L13}

\bibitem[\protect\citeauthoryear{{van den Bosch}}{{van den
  Bosch}}{2016}]{vandenbosch2016}
{van den Bosch} R. C.~E.,  2016, \mn@doi [\apj] {10.3847/0004-637X/831/2/134},
  831, 134

\makeatother
\end{thebibliography}

%\begin{thebibliography}{99}
%\end{thebibliography}

%%%%%%%%%%%%%%%%%%%%%%%%%%%%%%%%%%%%%%%%%%%%%%%%%%

%%%%%%%%%%%%%%%%% APPENDICES %%%%%%%%%%%%%%%%%%%%%

%\appendix

%%%%%%%%%%%%%%%%%%%%%%%%%%%%%%%%%%%%%%%%%%%%%%%%%%

% Don't change these lines
\bsp	% typesetting comment
\label{lastpage}
\end{document}